\documentclass[aps,prb,preprint,groupedaddress]{revtex4}
\usepackage{bm}
\usepackage{graphicx}

\newcommand{\lm}{\lambda}
\newcommand{\sg}{\sigma}
\newcommand{\bOm}{\bm{\Omega}}
\newcommand{\bx}{\bm{x}}
\newcommand{\br}{\bm{r}}
\newcommand{\bs}{\bm{s}}
\newcommand{\be}{\bm{e}}
\newcommand{\bA}{\bm{A}}
\newcommand{\bB}{\bm{B}}
\newcommand{\bR}{\bm{R}}
\newcommand{\bX}{\bm{X}}
\newcommand{\cA}{{\cal A}}
\newcommand{\cB}{{\cal B}}
\newcommand{\cS}{{\cal S}}

\begin{document}
\renewcommand\floatpagefraction{.001}
\makeatletter
\setlength\@fpsep{\textheight}
\makeatother

\setcounter{totalnumber}{1}
\title{The  directional contact distance of two ellipsoids:
Coarse-grained potentials for anisotropic interactions}
		
\author{Leonid Paramonov and S.N.~Yaliraki}
\email[]{s.yaliraki@imperial.ac.uk}
\affiliation{ Department of Chemistry, Imperial College London, South Kensington Campus, London SW7 2AZ, UK}

\date{\today}

\begin{abstract}

 We obtain  the 
distance of closest approach of the surfaces of    two arbitrary ellipsoids  valid at any orientation and separation,  measured along their 
inter-center vector. This directional distance is derived from the  Elliptic Contact Function. 
The geometric meaning behind this approach  is clarified.  An elliptic pair potential   for modelling  
arbitrary  mixtures of   elliptic particles, whether hard or soft,  is proposed based on this distance.
Comparisons to  Gay-Berne  potentials 
are discussed.  Analytic expressions for the forces and torques acting on the elliptic
particles are given.  

\end{abstract}

\maketitle

\section{Introduction}

Self-assembling systems of current interest in such diverse areas as molecular electronic materials to 
biological systems often involve molecular units or supra-molecular structures that are highly anisotropic in shape\cite{stupp}.
Typical examples range from lipids in biological membranes~\cite{frank}, to alkanethiols in 
self-assembled monolayers~\cite{goddard},  to carbon nanotubes and inorganic nanorods~\cite{Ratner03}. 
In the interplay between accuracy, and, simplicity and 
computational efficiency, such fairly rigid units are often approximated  as ellipsoids. 

The importance of the geometrical anisotropy in pair-wise interactions 
has been recognised early on. Berne and Pechukas introduced the Gaussian Overlap Potential (GOP) \cite{GOP} 
whose  generalisation  led to the widely used Gay-Berne potential (GB) \cite{GB}.   
The main idea behind this approach is the representation of a pair of particles  by their joint  elliptically stretched 
Gaussian distribution centered around the molecular centroids. The GB potential has been extensively  used for the modelling of the phase behaviour of  liquid crystals~\cite{Luckhurst93}. There is renewed interest in the possibility to study  dynamical behaviour, as for example in    
lipids in biological membranes~\cite{Voth2001}, and in  side-chains in coarse-grained protein force-fields~\cite{Scheraga97}.  
Part of the appeal of the form of the GB potential is that it allows for analytic derivation of forces and torques \cite{GBForces} acting on the particles 
which simplifies the modelling. The drawbacks are its lack of generality: for example, in its original form, it is only applicable to identical uniaxial elliptic particles, although it has been recently  
extended for the special case  of particles with biaxial symmetry \cite{GBE}.  
Another extension of the GB potential was made for an "ellipsoid in the sea of spheres" scenario \cite{EllSph},
where the main semi-axis of an elliptic particle is much bigger than the radius of the spherical particles interacting with it. However, as the geometry of the particles vary, so do the number and value of the parameters introduced in the potentials.

A different approach to this problem relies on the Elliptic Contact Function (ECF) introduced by Perram 
and coworkers \cite{John1,John2} which emphasises its geometrical aspects. The ECF approach calculates the distance of closest approach of two ellipsoids with given orientations. It gives the correct position of contact when the two ellipsoids are in tangent contact. Therefore it is often referred to as the "hard ellipsoids" approach. It exploits the algebra behind the representation of ellipsoids as quadratic forms 
to write the problem as an optimisation task that can be  solved efficiently. 
The GOP potential is known to be closely related to the Elliptic Contact Potential (ECP) derived from the ECF \cite{John2,GBE,Review93}.

Although applicable to any mixture, the ECF  has not yet been widely used in applications. 
The reasons may be due to  the lack of simplicity of the ECF approach 
and the lack of clarity of its geometric interpretation. Furthermore, the  potential derived from this approach  has several drawbacks: it  does not distinguish in energy between different relative orientations, it does not become isotropic at large separations and it artificially keeps the elliptic shape of the potential along the longer semi-axes of the ellipsoids.   
In this work, we clarify the geometric meaning of the ECF approach and the parameters associated with it. We  show  how the true distance of closest approach of the surfaces of two ellipsoids can be approximated well  directly from the ECF by the  directional distance of  closest approach  along their inter-center direction $R$. 
This allows us to develop a new type of elliptic potential  applicable for mixtures of  ellipsoids and/or spheres. 
We show that this can be done for any size, any orientation, and "hard" or "soft" particles. 
In all cases, the potential behaves isotropically at infinite separations and 
addresses the drawbacks discussed above.  
A  comparison with the GB potentials is also presented. 
Finally, we derive analytical expressions for the appropriate forces and torques that make Molecular Dynamics (MD) simulations  possible.

In Section~\ref{preliminaries} we give the preliminaries, namely review the main aspects  of the ECF  from the literature, 
in Section~\ref{geometry} we show how to obtain the directional distance of closest approach of two ellipsoids by the value of the ECF and compare it to the GB potentials.   
In Section~\ref{lennardjones} we show how it leads to a  LJ-type of  elliptic potential and 
in Section~\ref{forces} we derive analytic expressions for the  forces and torques acting on the particles due to 
the suggested potential.  Finally, 
in Section~\ref{discussion} we discuss the implications of this approach and comment on its advantages and  drawbacks.

\section{Preliminaries: the Elliptic Contact Function (ECF)}
\label{preliminaries}

\begin{figure}[htbp!]
\centering
\includegraphics{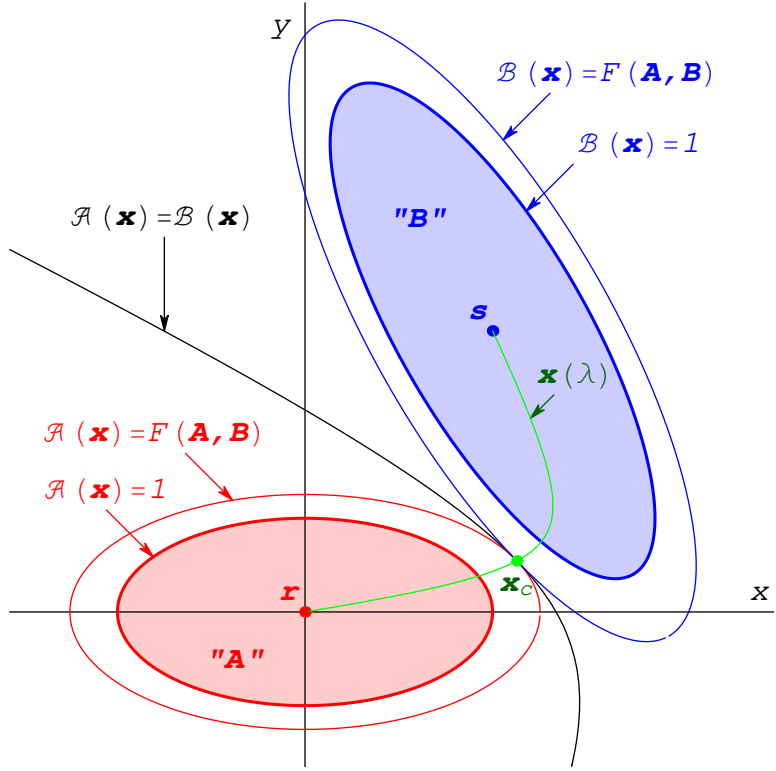}%{FigPlot10}
\caption{
The ellipsoid particles   "A" and "B", given by $\cA(\bx)=1$ and $\cB(\bx)=1$, are centered on $\br$ and $\bs$ respectively. We consider the local
reference frame attached to the centre  of  ellipsoid "A" aligned along its semi-axis.  We illustrate here, without loss of generality, the concepts behind the ECF approach  in the 2D case.  The original ellipsoids  are scaled up
(or down) by $\sqrt{F(A,B)}$ until they touch each other tangentially at the contact point $\bx_c$. The contact point $\bx_c$ is the intersection of the curve $\bx(\lm)$ (green line)  with the surface 
$\cA(\bx)=\cB(\bx)$ (black line).
The value of the ECF is illustrated by the scaled ellipsoids $\cA(\bx_c)=F(\bA,\bB)$ and
$\cB(\bx_c)=F(\bA,\bB)$.}
\label{TwoScaledElls}
\label{TwoEllsAndXC}
\end{figure}

The shapes of two anisotropic particles "A" and "B" centred at points $\br$ and $\bs$ are represented by the ellipsoids 
$\cA(\bx)=1$ and $\cB(\bx)=1$, respectively 
(Fig. \ref{TwoEllsAndXC}). 
$\cA(\bx)$ and $\cB(\bx)$ are quadratic forms given by 
$$\cA(\bx) = (\bx - \br) \cdot \bA \cdot (\bx - \br)$$ 
and 
$$\cB(\bx) = (\bx - \bs) \cdot \bB \cdot (\bx - \bs) .$$
The matrices  $\bA$ and $\bB$ can be expressed as   
\begin{equation}
\bA = \sum_{i=1,2,3} a_i^{-2} \bm{u}_i \otimes \bm{u}_i, \quad
\bB = \sum_{i=1,2,3} b_i^{-2} \bm{v}_i \otimes \bm{v}_i  \, ,
\label{ABOtimes}
\end{equation}
where $\bm{u}_i$ are the unit orthogonal vectors along the three semi-axes of ellipsoid "A" with lengths $a_i$ 
and, $\bm{v}_i$ are the unit orthogonal vectors along the semi-axes of ellipsoid "B" with lengths $b_i$.  The symbol "$\otimes$" in (\ref{ABOtimes}) is used to define the matrices $\bA$ and $\bB$ in dyadic form. 
 
The ECF is a measure of proximity of two ellipsoids,  
originally presented in \cite{John2} and subsequently reformulated in \cite{JohnUnPblshd}. 
The equivalence of the two definitions is shown in Appendix A. 
We begin with the latter formulation because it allows us to obtain a clear geometrical interpretation.  
To this end, we  define $\cS(\bx,\lm)$ as an affine combination of quadratic forms $\cA(\bx)$ and $\cB(\bx)$, 
\begin{equation}
\label{Fxl}
\cS(\bx, \lm) = \lm \cA(\bx) + (1-\lm) \cB(\bx) \, ,
\end{equation}
where $\lm$ is a parameter from the interval $[0,1]$. The ECF is defined as a solution to the following optimisation problem:
\begin{equation}
\label{sMinAB}
F(\bA,\bB) = \mathop{\it max}_{\lm} \mathop{\it min}_{\bx} \cS(\bx,\lm).
\end{equation}
The minimum $\bx(\lm)$ of  $\cS(\bx, \lm)$ for each value of the parameter 
$\lm$ can be found from 
\begin{equation}
\label{nablaF}
\nabla \cS (\bx, \lm) = 2 \left\lbrace 
	\lm \bA \cdot (\bx - \br) +
	(1-\lm) \bB \cdot (\bx - \bs)
\right\rbrace 
= 0
\end{equation}
%{\bf (*the previous equation got "=0" in the end*)}
as: 
\begin{equation}
\label{Xl}
\bx (\lm) = 
\left\lbrace  \lm \bA + (1-\lm) \bB \right\rbrace^{-1} \cdot
	\left\lbrace
		\lm \bA \cdot \br +
		(1- \lm) \bB \cdot \bs
	\right\rbrace .
\end{equation}
As a result, the optimisation problem given by Eq.\ref{sMinAB} is simplified to an unconstrained maximisation of the 
scalar function $\cS(\bx(\lm), \lm)$ by the scalar parameter $\lm$:
\begin{equation}
\label{ECF_Val}
F(\bA,\bB) = \mathop{\it max}_{\lm} \cS(\bx (\lm), \lm).
\end{equation}
Note that the curve $\bx(\lm): \lm \in [0,1]$ with $\bx(0)=\bs$ and $\bx(1)=\br$ connects the centres $\br$ and $\bs$ of the two ellipsoids, as can be seen in Fig \ref{TwoEllsAndXC}. Along this curve, the gradient vectors $\nabla \cA(\bx)$ and $\nabla \cB(\bx)$ are parallel.
%As follows from equation (\ref{nablaF}) at the point $\bx(\lm)$, the affine combination of these 
%gradient vectors of the quadratic forms are equal for all values of the parameter $\lm$. 
We will find later the following notation useful 
\begin{eqnarray}
&
\lm \nabla \cA(\bx(\lm)) =
2 \lm \bA \cdot (\bx(\lm)-\br) = 
\bX(\lm) \, ,
\label{BigXl}
&
\\
&
(1-\lm) \nabla \cB(\bx(\lm)) =
2 (1-\lm) \bB \cdot (\bx(\lm)-\bs) = 
-\bX(\lm).
\nonumber
&
\end{eqnarray}
%{\bf (* the last equation is splitted into two*)}

Earlier, it was shown (\cite{John1}) that the function $\cS(\bx(\lm), \lm)$ has a unique maximum
on the interval $\lm \in (0,1)$.
The value $\lm_c$ at which the function  reaches its maximum  is called
the {\it contact parameter} and the point $\bx_c=\bx(\lm_c)$ is called the {\it contact point}.
The derivative of the function  $\cS(\bx(\lm), \lm)$ with respect to $\lm$ is
\begin{eqnarray}
& \cS'(\bx(\lm), \lm) = 
\left\lbrace
\cA(\bx(\lm) )-\cB(\bx(\lm))
\right\rbrace + 
\nonumber
\\
& 2 \bx'(\lm)^T \cdot
\left\lbrace 
\lm 
\bA \cdot (\bx(\lm) - \br) +
(1- \lm) 
\bB \cdot (\bx(\lm) - \bs)
\right\rbrace .
\label{dFdl}
\end{eqnarray}
The term in the second curly braces vanishes along the curve $\bx(\lm)$ due to  Eq. (\ref{nablaF}).
Hence, at the extremum,
\begin{eqnarray}
& \cS'(\bx_c, \lm_c) = 
\cA(\bx_c )-\cB(\bx_c) = 0 \, ,
\label{dFdlsmall}
\end{eqnarray}
from which we can see  that the contact point $\bx_c$ is the intersection of the curve $\bx(\lm)$ 
with the surface $\cA(\bx)=\cB(\bx)$ (Fig. \ref{TwoEllsAndXC}). 
Eq.\ref{dFdlsmall} leads to 
\begin{equation}
\label{AeqB}
\cA (\bx_c) = \cB (\bx_c).
\end{equation} 
Substitution of the last equation back into Eq. (\ref{Fxl}) gives the following simple 
interpretation of the ECF value
\begin{equation}
\label{XIeq}
F(\bA,\bB) = \cA (\bx_c) = \cB (\bx_c).
\end{equation}

The contact point $\bx_c$ lies on the ellipsoid $\cA(\bx)=F(\bA,\bB)$ due to Eq. (\ref{XIeq}).
Similarly for $\cB(\bx)$ the contact point $\bx_c$ lies on the ellipsoid $\cB(\bx)=F(\bA,\bB)$.
The two ellipsoids are in tangent contact (Fig. \ref{TwoScaledElls}) due to Eq. (\ref{nablaF}).
As a result, the value of the ECF serves as a criterion of approach of any two elliptic particles.
If the value $F(\bA,\bB)$ is below unity, then the original ellipsoids $\cA(\bx)=1$ and
$\cB(\bx)=1$ overlap and vice versa.
When $F(\bA,\bB)$ is equal to  one, the original ellipsoids "A" and "B" are in tangent contact.

The contact parameter $\lm_c$ can be found numerically as a solution of Eq. (\ref{AeqB}) by
approaching  the surface $\cA (\bx) = \cB (\bx)$ along the curve $\bx(\lm)$: 
\begin{equation}
\label{lc_EQ}
\lm_c: \cA(\bx(\lm_c) )-\cB(\bx(\lm_c))=0.
\end{equation}
The convergence of this iterative process is usually fast. 
%%{\bf Numerical aspects of this iterative process as well as ways to speed it up will be discussed elsewhere.} 
Numerical aspects of this iterative process as well as ways to speed it up will be discussed elsewhere.

The value of the ECF can be further expressed in the form $F(\bA,\bB) = R^2 \, f(\bA,\bB)$, 
where $R$ is the inter-center distance between particles "A" and "B" (See Appendix A) and $f$ depends only on orientation, which leads to the definition of  
the Perram-Wertheim (PW) range parameter $\sg_{PW}(\bA,\bB)$ (\cite{GBE}) as:
\begin{equation}
\label{sPW}
\sg_{PW}(\bA,\bB) = \frac{R}{\sqrt{F(\bA,\bB)}} = \frac{1}{\sqrt{f(\bA,\bB)}}.
\end{equation} 
There is a well-known close relationship \cite{John2,GBE} between this parameter and the  
Berne-Pechukas (BP) range parameter, $\sg_{BP} (\bA,\bB)$, first introduced in \cite{GOP} for the GOP.
$\sg_{BP} (\bA,\bB)$ can be expressed in terms of  
$\cS(\bx(\lm), \lm)$ (see Appendix B) by substituting $\lm_c=1/2$,
\begin{equation}
\label{sGOP}
\sg_{BP} (\bA,\bB) = \frac{R} { \sqrt{\cS(\bx(1/2),1/2)} }.
\end{equation} The BP range parameter is sometimes considered as a mean value approximation of the PW range parameter~\cite{GBE,GBE1}, where $\lm_c$ is approximated by $1/2$ (see Eq.~\ref{sGOP}). 
Additionally, Eq. \ref{sPW} is often treated  as the distance of 
closest approach of two ellipsoids, therefore Eq.~(\ref{sGOP}) is believed to be an approximation of the distance of closest approach (Eq.~\ref{dMinAB}). We will return to this point in Sec.~\ref{comparison}.

\section{Approximations and geometrical interpretations of the distance of closest approach of two ellipsoids}
\label{geometry}

We  now show how to obtain  the distance of closest approach of the surfaces of two ellipsoids using the ECF value.
The true distance of closest approach, $d$, is a solution of the following minimisation problem 
\begin{equation}
\label{dMinAB}
d(\bA,\bB) = \mathop{\it min}_{\tilde{\bx}_a,\tilde{\bx}_{b}} ||\tilde{\bx}_{a} - \tilde{\bx}_{b}||,
\end{equation}
subject to two constraints
\begin{equation}
\label{ConstrMinAB}
\cA(\tilde{\bx}_a)=1, \quad \cB(\tilde{\bx}_b)=1, 
\end{equation}
namely, that  $\tilde{\bx}_a$ and $\tilde{\bx}_b$ are points on the surface of ellipsoids ``A'' and ``B'', respectively.

 We will now show how to approach this distance using the ECF value,   both  from above  using the geometrical properties of the ECF, and from below,  using a mechanical analogy.
%%%%%%%%%%%%Above
\subsection{The directional contact distance, $d_R$}
\label{dRsection}
\begin{figure}[htbp!]
\centering
\includegraphics{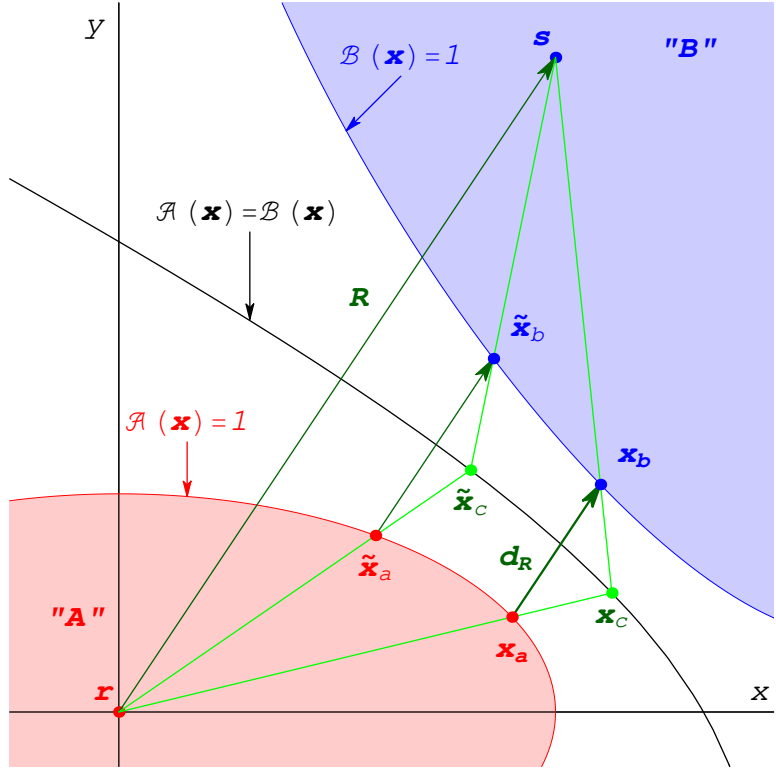}%{FigPlot10}
\caption{The  directional distance of closest approach, $d_R(A,B)$,  is the minimum distance  between the surfaces of two ellipsoids   that is parallel to the inter-center vector R. It  is given by the  sub-contact points $\bx_a$ and $\bx_b$, which are on  the surface of ellipsoids ``A'' and ``B'', respectively. It is a good approximation from above to the true contact distance.    
The value of the PW 
approach parameter $\sg_{PW}(\bA,\bB)$
can be  understood by the  geometry of the triangle of points $\br$, $\bs$ and $\bx_c$, as the length of the vector $\bR -(\bx_a$- $\bx_b)$ as soon as $(\bx_a$- $\bx_b)$ becomes parallel to $\bR$ }
\label{ExtraSubPoints}
\end{figure}

An intersection of the line segment between the contact point $\bx_c$ and the center $\br$  
with the surface of the ellipsoid $\cA(\bx)=1$ (Fig.~\ref{ExtraSubPoints}) can be found 
using the value $F(\bA,\bB)$ as: 
\begin{equation}
\bx_a = \br+(\bx_c-\br) F(\bA,\bB)^{-1/2}.
\label{xa}
\end{equation}
The point $\bx_a$ will be  called {\it a sub-contact point of A}. 
The {\it sub-contact point of B}, $x_b$ can be found in the same way, namely as the intersection of the line segment 
between points $\bx_c$ and $\bs$ with the surface of the ellipsoid 
$\cB(\bx)=1$
\begin{equation}
\bx_b = \bs+(\bx_c-\bs) F(\bA,\bB)^{-1/2}
\label{xb}.
\end{equation}
The vector $(\bx_b-\bx_a)$ between the sub-contact points of A and B is parallel to the inter-center vector $\bR$ and is given by 
\begin{eqnarray}
&
(\bx_b-\bx_a) = (\bs-\br) (1-F(\bA,\bB)^{-1/2}) = 
\hat{\bR} \, d_R(\bA,\bB) \, ,
&
\end{eqnarray} 
where 
\begin{equation}
\label{dR}
d_R(\bA,\bB)=R(1-F(\bA,\bB)^{-1/2})= R-\sg_{PW}(\bA,\bB). 
\end{equation}
The geometrical meaning of the PW range parameter becomes now clear from  Eq. (\ref{dR}): it is the sum of projections 
of vectors $(\bx_a-\br)$ and $(\bs-\bx_b)$ on the inter-particle vector $\bR$ or simply the length of the
vector $\bR-(\bx_b-\bx_a)$ as soon as $(\bx_b-\bx_a)$ is parallel to $\bR$
(Fig. (\ref{ExtraSubPoints})). The distance $d_R$ (Eq. \ref{dR}) has a meaning of the shortest directional distance between two ellipsoids 
measured along the direction of their inter-particle vector $\bR$. 

To see this note that on the surface $\cA(\bx) = \cB(\bx)$ any affine combination $\cS(\bx,\lm)$ with 
any value of the parameter $\lm$ is equal to both $\cA(\bx)$ and $\cB(\bx)$. 
At the same time for $\lm=\lm_c$ we have shown, that $\cS(\bx,\lm_c)$ reaches it's 
minimum at the contact point $\bx_c$. As a result, the optimisation problem (\ref{sMinAB})-(\ref{Fxl}) 
can be reformulated as a problem of finding  the minimum of $\cA(\bx)$ or $\cB(\bx)$ 
on the surface $\cA(\bx)=\cB(\bx)$. 
%Except the convergence of the problem (\ref{Fxl}) is much faster.

Note that Eqs. (\ref{xa}) and (\ref{xb}) hold not only for the contact point $\bx_c$ 
but for any point $\tilde{\bx}_c$ (see Fig. \ref{ExtraSubPoints}) 
on the surface $\cA(\bx) - \cB(\bx) = 0$ 
in the following way:
\begin{equation}
\tilde{\bx}_a = \br+(\tilde{\bx}_c-\br) \cS(\tilde{\bx}_c)^{-1/2} \, ,
\label{xA}
\end{equation}
\begin{equation}
\tilde{\bx}_b = \bs+(\tilde{\bx}_c-\bs) \cS(\tilde{\bx}_c)^{-1/2} \, ,
\label{xB}
\end{equation}
where $\cS(\bx)$ means here any  affine combination of $\cA(\bx)$ and $\cB(\bx)$. The resulting vector $(\tilde{\bx}_b-\tilde{\bx}_a)$ is also parallel to the inter-particle vector $\bR$ 
(Fig. \ref{ExtraSubPoints}).
\begin{equation}
(\tilde{\bx}_b-\tilde{\bx}_b) = \bR (1-\cS(\tilde{\bx}_c)^{-1/2}).
\end{equation} The minimum of the value $\cS(\tilde{\bx}_c)$ is  $F(\bA,\bB)$. It is easy to show, that the value 
$(1-\cS(\tilde{\bx}_c)^{-1/2})$ reaches its minimum together with $\cS(\tilde{\bx}_c)$. As a result, the distance $d_R(\bA,\bB)$ has a meaning of minimum length of the vector $(\tilde{\bx}_b-\tilde{\bx}_a)$ 
parallel to $\bR$.  

If we consider the value $d_R(\bA,\bB)$ 
as a minimum distance along any arbitrary direction independently from the 
inter-particle vector, then its minimum among all possible directions naturally is the distance of closest approach of 
two ellipsoids $d(\bA,\bB)$. As a result the following inequality holds while $d(\bA,\bB)$ is greater than  zero:
\begin{equation}
d(\bA,\bB) \leq d_R(\bA,\bB) \, .
\label{ddR}
\end{equation}
%%%Below
\subsection{The distance from below, $d_n$}
An estimation of the distance of Eq. (\ref{dMinAB}) from below is given by the distance 
between two parallel planes $\alpha_a$ and $\alpha_b$, which are tangent to ellipsoids "A" and "B" at the  sub-contact points 
$\bx_a$ and $\bx_b$ (Fig. \ref{Planes}):
\begin{equation}
d_{\bm{n}}(\bA,\bB) = 
\hat{\bX}_c \cdot (\bx_b-\bx_a) = 
\hat{\bX}_c \cdot \bR \, (1-F(\bA,\bB)^{-1/2}) \, ,
\label{dn}
\end{equation}
where $\hat{\bX}_c$ is the unit vector in the direction of the gradient vector (Eq. \ref{BigXl}) 
$$
\hat{\bX}_c = 
\frac{\bX_c}{||\bX_c||} = 
\frac{\nabla \cA(\bx_a)}{||\nabla \cA(\bx_a)||} = 
- \frac{\nabla \cB(\bx_b)}{||\nabla \cB(\bx_b)||}.
$$

In fact, the maximum of the distance (\ref{dn}) among all possible positions of points $\tilde{\bx}_A$ and $\tilde{\bx}_B$ 
on the surfaces of ellipsoids "A" and "B", such that
the %$\nabla\cA(\bx_A)$ and $\nabla\cB(\bx_B)$ 
tangent planes $\alpha_a$ and $\alpha_b$ are parallel, is equal to the true contact distance  $d(\bA,\bB)$ 
as long as the ellipsoids "A" and "B" are not overlapping. 

\begin{figure}[htbp]
\centering
\includegraphics{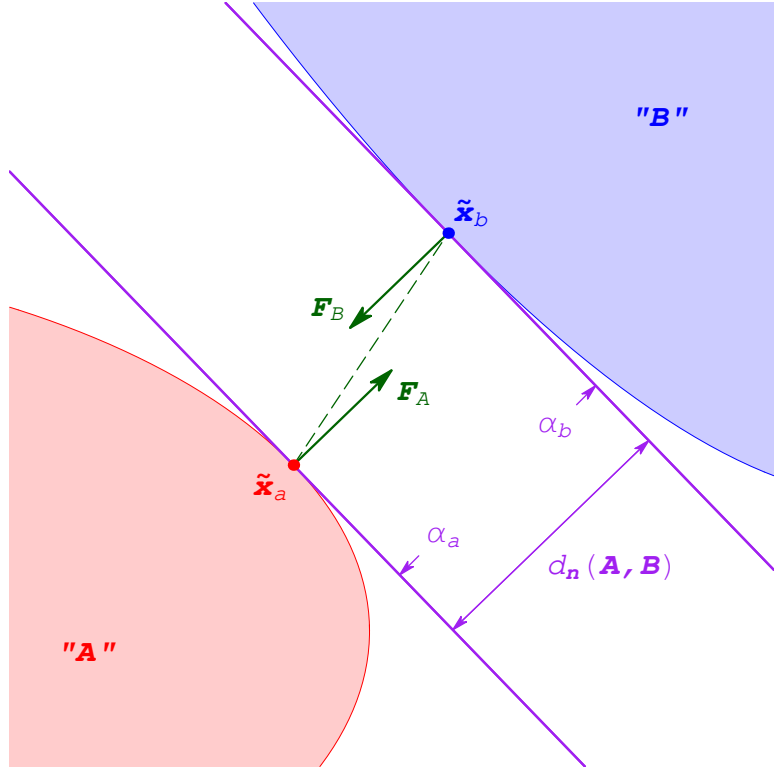}%{FigPlot12}
\caption{ The distance $d_n(A,B)$  is defined as the distance between two parallel planes $\alpha_a$ and $\alpha_b$ which are tangent  at any point $\tilde{\bx_a}$ and $\tilde{\bx_b}$ on the surface of  ellipsoids ``A'' and ``B'', respectively. It approaches the true distance of closest approach  from below. The maximum distance $d_n(A,B)£$ coincides with  the true distance of closest approach. This can be understood from a mechanical point of view: if the planes are kept apart by  a constant force ${\bf F}$, then the equilibrium point reached   corresponds to the maximum distance between planes, which is also the distance of closest approach.}
\label{MaxDNDist}
\end{figure}

\begin{figure}[htbp]
\centering
\includegraphics{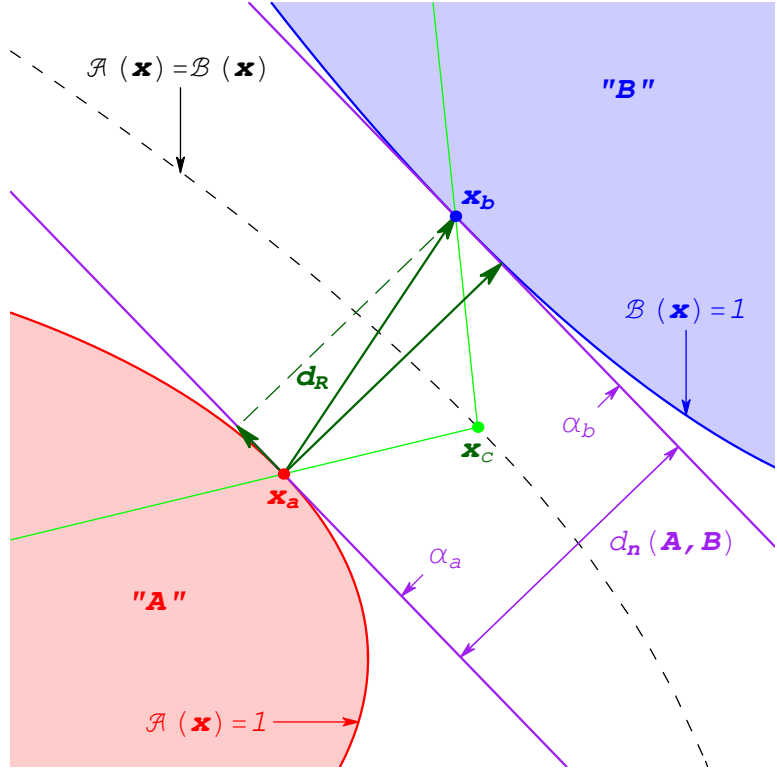}
\caption{A closer look at the sub-contact points $\bx_a$ and $\bx_b$, which define the directional distance of closest approach $d_R$. The planes $\alpha_a$ and 
$\alpha_b$ are  tangent to the ellipsoids $\cA(\bx)=1$ and $\cB(\bx)=1$ respectively at these points (shown with 
purple lines). The distance $d_n$ approaches  the true distance of closest approach  from below, while $d_R$ approaches it  from above.}
\label{Planes}
\end{figure}

To show this, we  consider a mechanical analogy. We consider the  parallel planes $\alpha_a$ and $\alpha_b$  between two fixed rigid convex bodies 
"A" and "B" which are separated by some distance and  kept apart by a constant force $\bm{F}$.  The two planes will move away from each other until they touch the bodies "A" and "B" at  points $\tilde{\bx}_a$ and $\tilde{\bx}_b$ respectively  
(see Fig. \ref{MaxDNDist}). The reaction forces $\bm{F}_A$ and $\bm{F}_B$ from the bodies "A" and "B" will act on the 
planes $\alpha_a$ and $\alpha_b$ along normal vectors of the surfaces of "A" and "B" at the points 
$\tilde{\bx}_a$ and $\tilde{\bx}_b$. 
The normal vectors at the surfaces of "A" and "B" at $\tilde{\bx}_a$ and $\tilde{\bx}_b$ are parallel due to the fact that $\alpha_a$ and $\alpha_b$ are parallel themselves. 

The stationary/equilibrium point will be reached when the total force $\bm{F}_A+\bm{F}_B$ and the total torque 
$(\bx_B-\bx_A) \times \bm{F}_B$ acting on the system of the planes are both equal to zero. The stationary point will 
correspond to the maximum possible distance between the planes because of the force $\bm{F}$ that is acting  to separate them. The total force is always equal to zero $\bm{F}_A+\bm{F}_B=0$ because $\bm{F}_A=\bm{F}$ and $\bm{F}_B=-\bm{F}$. 
The total torque can be equal to zero only if $(\tilde{\bx}_a-\tilde{\bx}_b)$ is parallel to the normal vectors of the surfaces 
of "A" and "B" at $\tilde{\bx}_a$ and $\tilde{\bx}_b$, which is a solution of the optimisation problem for the distance of 
closest approach $d(\bA,\bB)$ (Eq. (\ref{dMinAB})) as well. As a result the maximum distance between planes $\alpha_a$ and $\alpha_b$ trapped between two convex bodies is equal 
to the distance of closest approach of the surfaces of these two bodies. 

This leads to 
%The inequality $d_n \le d_R$ holds always because the $d_n$ is the scalar product of the unit vector $\hat{\bX}_c$ 
%with vector $(\bx_b-\bx_a)$. The same implies that $d_n=0$ if $d_R=0$.
%%%%%%%%%%%%%%%%%%%%%%%%%%%%%%%%%%%%%%%%%%%%%5
\begin{equation}
\label{TheBigUnEq}
0 < d_{\bm{n}}(\bA,\bB) \leq d(\bA,\bB) \leq d_R(\bA,\bB) \, ,
\end{equation}
while $d(\bA,\bB) > 0$.  
All three values $d_{\bm{n}}(\bA,\bB)$, $d(\bA,\bB)$ and $d_R(\bA,\bB)$ are equal to zero, 
when the ellipsoids are in tangent contact
\begin{equation}
\label{TheBigUnEq1}
0 = d_{\bm{n}}(\bA,\bB) = d(\bA,\bB) = d_R(\bA,\bB).
\end{equation}
The inequality (\ref{TheBigUnEq}) does not hold  when the two ellipsoids overlap. The distance $d(\bA,\bB)$, the 
solution of the optimisation task (\ref{dMinAB})-(\ref{ConstrMinAB}), remains zero whenever "A" and "B" overlap. However,  the values $d_R(\bA,\bB)$ and $d_{\bm{n}}(\bA,\bB)$ are below zero and characterise 
the overlap of the ellipsoids. As a result, the distance $d_R(\bA,\bB)$ can be used in both "soft" 
and "hard" potentials. 

In practise,  $d_{\bm{n}}(\bA,\bB)$ is  closer to the true distance $d(\bA,\bB)$ for small separations while    $d_R(\bA,\bB)$ is a very good approximation at larger separations and behaves isotropically for infinite separations. 
%additional scalar product of the vector $(\bx_b-\bx_a)$ with the vector $\hat{\bX}$. 
Even though the distance $d(\bA,\bB)$ (\ref{dMinAB}) can be used to build a pair potential 
between two elliptic particles \cite{EveraersEjtehadi2004},
the usage of a pair potential based on this distance in molecular simulations is computationally 
expensive because it requires a solution of the problem (\ref{dMinAB})-(\ref{ConstrMinAB})
at least once at  each integration step for each pair of particles. Although the calculation of the ECF as a proximity measure still involves the solution of the optimisation task 
(\ref{Xl})-(\ref{ECF_Val}) for each pair of particles 
on each simulation step, the resulting optimisation problem is simpler and computationally more efficient than the problem
(\ref{dMinAB})-(\ref{ConstrMinAB}). 
Additionally, we will show that the usage of the  range parameter of Eq.~(\ref{sPW}) leads to simple and compact expressions for 
forces and torques acting on the particles. 

\subsection{On the relation between the BP  and PW range parameters}
\label{comparison}

%In this section we will show  that usage of the BP range parameter $\sg_{BP}(\bA,\bB)$ 
%instead of the PW range parameter $\sg_{PW}(\bA,\bB)$ can, in the absence of symmetry, lead to 
%deviations of volume and shape of the elliptic particles, which may act as 
%artificial "ordering" forces and overselect for ordered configurations. 
%{\bf (*I'm not sure about last sentence. Do we need "..., in the absence of symmetry,..." here? *)}

We can now return to the relation of  the BP range parameter  as a mean value approximation 
to  the PW range parameter (\cite{EveraersEjtehadi2004,GBE}). 
To examine this further, let's consider the value of  $\cS(\bx,\lm)$ 
on the curve $\bx(\lm)$ for different relative orientations. We first consider  two identical uniaxial elliptic particles. 
The   GOP and  GB potentials were originally introduced for this kind of particles~\cite{GOP,GB}. We will show that  the function $\cS(\bx(\lm),\lm)$ becomes symmetric on the interval 
$\lm \in \left[ 0, 1 \right]$ whenever there is a point or a line or a plane of symmetry between the two ellipsoids. As a result the maximum $F(\bA,\bB)$ of the function $\cS(\bx(\lm),\lm)$ can only occur at $\lm_c=1/2$. Hence, when the quadratic forms $\cA(\bx)$ and $\cB(\bx)$ are symmetric,  the BP  range parameter $\sg_{BP}(\bA,\bB)$ agrees with  the value  of the  PW  range parameter $\sg_{PW}(\bA,\bB)$. However,  for asymmetric configurations, this  is generally not true. 

To proceed we express  the shape matrix of a spheroid  as 
\begin{equation}
\bA= l^{-2} \, \bm{u} \otimes \bm{u} + d^{-2} \, (\bm{E}-\bm{u} \otimes \bm{u}) \, , 
\label{spheroidMatrix} 
\end{equation}
where 
$2 l$ is the length of the ellipsoid, 
$2 d$  its breadth, 
$\bm{u}$  a unit vector of the the main axis
and $\bm{E}$  a unit tensor given by  $\bm{E} = \sum\limits_{j=1}^3 \be_j \otimes \be_j.$
%$$
%\bm{E} = \sum\limits_{j=1}^3 \be_j \otimes \be_j
%$$
Let's for simplicity move the origin of the reference frame into a centre of symmetry $\bm{c}$
 (this point is not necessarily unique for different types of symmetry). 
Vectors designating directions and positions in this reference frame are  denoted by Greek letters.  A vector $\bm{\xi}$ is transformed into its symmetric image $\tilde{\bm{\xi}}$ by a matrix $\bOm$ which does not change its length but changes its directions:
\begin{equation}
\tilde{\bm{\xi}} = \bOm \cdot \bm{\xi} \, .
\label{eqOmegaXi}
\end{equation} 
For the matrix $\bOm$ to be a symmetric transformation, we require the following property:
\begin{equation}
\bOm \cdot \bOm =\bm{E} \, ,
\end{equation} 
which implies that the determinant of $\bOm$ can be $1$ or $-1$.

The quadratic forms $\cA(\bx)$ and $\cB(\bx)$ are symmetric 
with respect to the centre of symmetry $\bm{c}$ if the centre $\bs$ of  $\cB(\bx)$ is 
a symmetric image of the centre $\br$ of $\cA(\bx)$, so that 
\begin{equation}
\br=\bm{c}+\bm{\eta} \, , \quad
\bs=\bm{c}+\tilde{\bm{\eta}} \, , \quad 
\tilde{\bm{\eta}}=\bOm \cdot \bm{\eta}
\end{equation} 
,and, the matrix $\bB$ is a symmetric image of the matrix $\bA$, satisfying  
\begin{equation}
\bB=\bOm \cdot \bA \cdot \bOm.
\label{eqOmegaAB}
\end{equation} 
Eq.  (\ref{eqOmegaAB})  includes in addition the  symmetries of   particle "A".

The possible cases of symmetry of quadratic forms $\cA(\bx)$ and $\cB(\bx)$ are  
the point symmetry $\bOm=-\bm{E}$, the line symmetry, $\bOm=2 \, \bm{l} \otimes \bm{l}-\bm{E}$, and the plane symmetry $\bOm=\bm{E} - 2 \, \bm{n} \otimes \bm{n}$, where 
%\begin{itemize}
%\item the point symmetry $\bOm=-\bm{E}$;
%\item the line symmetry $\bOm=2 \, \bm{l} \otimes \bm{l}-\bm{E}$;
%\item 
%\end{itemize} 
$\bm{l}$ is a unit vector along the axis of symmetry and
$\bm{n}$ is a unit vector normal to the plane of symmetry. 
In the case of line symmetry, any point of the actual axis of symmetry can serve as a centre of symmetry $\bm{c}$, while 
in the case of plane symmetry, any point of the plane of symmetry can be considered as a centre of symmetry $\bm{c}$.

Using Eqs. (\ref{eqOmegaXi})-(\ref{eqOmegaAB}), it can be shown that for any point $\bm{\xi}$ and its image $\tilde{\bm{\xi}}$, 
the values of the symmetric quadratic forms satisfy the following equations:
\begin{equation}
\cA(\bm{\xi}) = \cB(\tilde{\bm{\xi}}), \quad 
\cA(\tilde{\bm{\xi}}) = \cB(\bm{\xi}) \, .
\end{equation} 

The  function $S$ of Eq \ref{Fxl} now  becomes 
\begin{eqnarray}
&
\label{eqSymS}
\cS(\bm{\xi},\lm) = 
\lm \cA(\bm{\xi}) +(1-\lm) \cB(\bm{\xi}) = 
&
\\
&
\lm \cB(\tilde{\bm{\xi}}) +(1-\lm) \cA(\tilde{\bm{\xi}}) = 
\cS(\tilde{\bm{\xi}},(1-\lm)) 
\nonumber
&
\end{eqnarray}
and its gradient is given by 
\begin{equation}
\nabla \cS(\tilde{\bm{\xi}},(1-\lm)) = \bOm \cdot \nabla \cS(\bm{\xi},\lm) \, .
\label{eqSymNS}
\end{equation}

From Eq. (\ref{eqSymNS}) it follows that, if a point $\bm{\xi}$ belongs to the curve  
$\bx(\lm)$  (Eq.~(\ref{Xl})), its symmetric image belongs to the curve  
$\bx(\lm)$ as well. Further from Eq. (\ref{eqSymS}) it follows that whenever a symmetry described by Eqs.
(\ref{eqOmegaXi})-(\ref{eqOmegaAB}) is present, the function $\cS(\bx(\lm),\lm)$ becomes symmetric 
on the interval $\lm \in [ 0, 1 ]$.  As a result, the maximum $F(\bA,\bB)$ of the function $\cS(\bx(\lm),\lm)$ can only occur at  $\lm_c=1/2$. This shows that 
the BP and PW parameters are equivalent for symmetric quadratic forms, namely for symmetric configurations of equivalent particles. 

If we now  inspect the zero level of the GB pair potential depending on the relative orientation of two particles 
\begin{equation}
\label{GB0}
U_{GB}(\bA,\bB) = 0 \, ,
\end{equation}
we can see that the level (\ref{GB0}) does not depend on the GB strength parameters (see Appendix B) and it is equivalent to 
\begin{equation}
\sg_{BP}(\bA,\bB) = 1 \, ,
\label{S05e1}
\end{equation}
which is in turn equivalent to
\begin{equation}
\cS(\bx(1/2),1/2) = 1 \, .
\label{S05func}
\end{equation}

\begin{figure}[htbp!]
\centering
\includegraphics{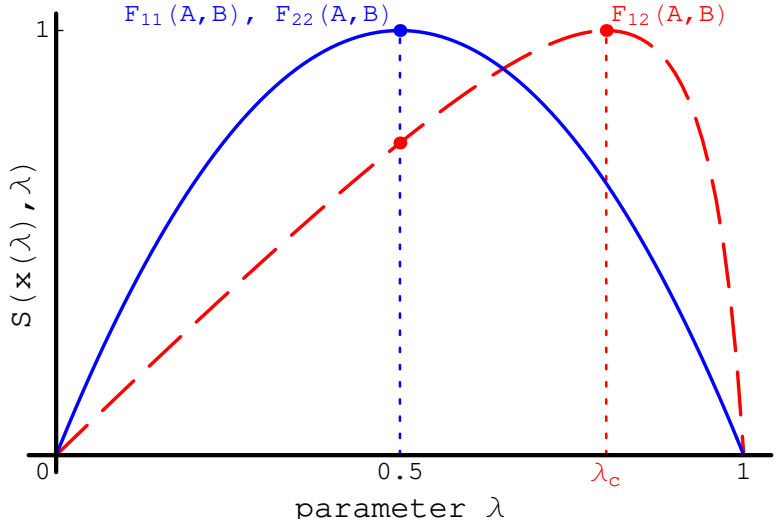}%{FigPlot11}
\caption{ $\cS(\bx(\lm),\lm)$ is shown as a function of the parameter $\lambda$  for the ``2-to-2'' ("side-to-side"),
``1-to-1'' ("end-to-end") and ``1-to-2'' ("side-to-end") relative orientations of two identical ellipsoids. The  ECF value, ${F_{ij}(A,B)}$, is the maximum of the function $S$ and gives the contact parameter $\lambda_c$ for  each orientation.  The approximation of 
$\lm_c$ by $1/2$ holds for symmetric configurations of identical ellipsoids such as ``2-to-2''  and ``1-to-1'',      but  fails in the asymmetric ``1-to-2''    configuration. This is expected to hold  for asymmetric configurations in general, which includes any configuration of non-identical particles.}
\label{SFunctionPlot}
\end{figure}

In the case of point symmetry the matrix $\bB$ is equal to the matrix $\bA$. 
\begin{equation}
\bB = 
\bOm \cdot \bA \cdot \bOm =
(-\bm{E}) \cdot \bA \cdot (-\bm{E}) =
\bA \, ,
\end{equation} 
which corresponds to "perfectly aligned" configurations including "side-to-side" and "end-to-end" configurations, which 
are often used for the adjustment of GOP and GB potentials. (In this work, we use the ``i-to-j''notation to refer to the different configurations. For example, ``end-to-end'' is denoted by ``1-to-1'', ``side-to-side'' by ``2-to-2'', ``side-to-end'' by  ``1-to-2''.)  As a result, the BP range parameter predicts the value of the 
PW range parameter correctly. 
However, the ``1-to-2'' orientation of the molecules is not symmetric and here  the approximation $\lm_c=1/2$ fails
(Fig. \ref{SFunctionPlot}). 
In fact, Eq. (\ref{S05e1}) will be satisfied
when the actual ellipsoids are already separated by some distance above zero  
due to the fact that in asymmetric configurations 
the value $\cS(\bx(1/2),1/2)$ is always an underestimation of the ECF value $F(\bA,\bB)$ 
(Fig. \ref{SFunctionPlot}).
%%{\bf (*"In fact...due to the fact..."*)}
This means that  the size of the molecule given by (\ref{GB0}) for the  "side-to-end" configuration 
(and any other configuration such that $\lm_c \neq 1/2$)
will be increased compared to a symmetric configuration (such as "side-to-side" and "end-to-end")
of the same molecule and the same potential. 
Such an increase of the volume of the particles might act as an artificial "ordering force", which will force elliptic 
particles into symmetric (ordered) configurations. 

%A further inspection shows, that of the shape of the level $U_{GB}=0$ in "side-to-end" configuration 
%does not correspond to interaction of two elliptic particles. 
%As a result the elliptic shape of the particles is not preserved.
%For elliptic particles with high eccentricity the resulting shape even become non-convex.
%(*This bit needs to be discussed deeper probably. It is a long discussion - how we define this resulting shapes. 
%Do you remember those figures  with ellipsoids, which where thrown away?
%Referees will stuck on this bit for sure. It might be thrown away for a while.*)

This analysis is of course far from  complete. The total pair potential also depends on the GOP strength parameter, so that the resulting deviations of the final shape of the particles are rather complex. 
In general, we expect  the described deviation to grow with the  ellipticity of the  interacting spheroids. 
Note, that for unequal elliptic particles, the approximation $\lm_c=1/2$ fails everywhere except for some rare cases.

\section{An elliptic potential based on the directional distance, $d_R$}
\label{lennardjones}

We  now use the directional distance $d_R(\bA,\bB)$ obtained in Sec.\ref{dRsection} to construct a new elliptic potential. Here, we concentrate on the LJ 12-6 form as an example of a radial potential. 

The LJ potential for two different spherical particles "A" and "B"  has the following form~\cite{liquidsbook}:
\begin{equation}
U_{LJ} (A,B) = 4 \epsilon_{AB} 
	\left[ 
		\left( 
			\frac{\sg_{AB}}{\rho_{AB}}
		\right)^{12} 
		-
		\left( 
			\frac{\sg_{AB}}{\rho_{AB}}
		\right)^6 
	\right] 
\, ,
\label{ULJ}
\end{equation}
where 
\begin{equation}
\sg_{AB} = (\sg_{A}+\sg_{B})/2, \quad
\epsilon_{AB} = \sqrt{\epsilon_{AA} \epsilon_{BB}} \, ,
\label{sigmAB}
\end{equation}
and $\rho_{AB}$ is the inter-center distance. 
Eq. (\ref{sigmAB}) represents the well known Lorentz-Berelot mixing rules 
for Van-der-Waals radii $\sg_{A}$ and $\sg_{B}$ of interacting particles and their depths parameters
$\epsilon_{AA}$ and $\epsilon_{BB}$ commonly used in the LJ potential. 
The  parameter $\epsilon_{\alpha\alpha}$ has the meaning of the depth of the 
potential well  for the  interaction of two identical particles 
of type "$\alpha$". 
%The (\ref{sigmAB}) gives the depth of the LJ potential well for interaction of two particles of different types. 

%The mixing rules (\ref{sigmAB}) are simple, but not always accurate enough.
%There are others more sophisticated mixing rules as well. 

The distance $d_R(\bA,\bB)$ can be consistently compared with the inter-center distance $R$. 
Using Eq. (\ref{dR}), the PW range parameter $\sg_{PW}(\bA,\bB)$
%$\sg_{PW}(\bA,\bB)=R-d_R(\bA,\bB)$
can be treated as the sum of radii of two elliptic particles at a given relative orientation. 
From this point of view, $\sg_{PW}(\bA,\bB)$ can be considered as an orientation-dependent
"mixing rule" for two elliptic particles similar to $\sg_{AB}$ in Eq.\ref{sigmAB}. 
Using this analogy, the following elliptic potential can be built in Lennard-Jones form: 
\begin{eqnarray}
&
U = 4 \epsilon_0 
\left[ 
	\left( 
	\frac{R \, F(\bA,\bB)^{-1/2}}{ R }
	\right) ^{12} -
	\left( 
	\frac{R \, F(\bA,\bB)^{-1/2}}{ R }
	\right) ^{6} 
\right] =
& \nonumber \\
& 
4 \epsilon_0 
\left[ 
F(\bA,\bB)^{-6} - F(\bA,\bB)^{-3}
\right]\, ,
&
\label{ECP}
\end{eqnarray}
which is  the Elliptic Contact Potential (ECP) developed by Perram and co-workers \cite{John1,John2}. 
We can now see that this potential is a consistent generalisation of the Lennard-Jones potential 
in the case of elliptic particles when one considers 
%the contact distance $d_R(\bA,\bB)$ as a generalisation of 
the inter-particle distance $R$ together with the "mixing rule" $\sg_{PW}(\bA,\bB)$.  
The value $\epsilon_0$ has the meaning of the potential minimum. 

However, the potential (\ref{ECP}) in this form has several disadvantages. First of all, it does not become isotropic 
at large separations of the particles. Secondly, the depth of the potential minimum remains the same 
for all relative orientations of the particles "A" and "B". 
And  lastly, the shape of the potential well is not realistic because it is wider along the longer semi-axes of the ellipsoids than the original potential. These disadvantages prevent the potential of Eq.
(\ref{ECP}) from being extensively used in 
applications. Here, we would like to build an extension to the ECP, which   
preserves the good features of the ECF approach but
%frees the potential 
frees it from the disadvantages mentioned above.

These problems come from the 
 "Lennard-Jones like" reduced distance $(R/\sg_{PW}(\bA,\bB))$ used in the ECP  (Eq. \ref{ECP}). This distance  becomes one when the ellipsoids are in contact and the potential goes to zero and  keeps an "elliptic shape" at large separations that  causes both the unrealistic anisotropy of the ECP at long distances 
and the unrealistic shape of the potential well. 
However, the directional distance $d_R(\bA,\bB)$ along the inter-particle vector  itself becomes isotropic at large separations. Using this distance instead, a "shifted form" of the potential (\ref{ECP}) can be built as:
\begin{eqnarray}
&
U (\bA,\bB)
= 4 \epsilon_0 
\left[ 
	\left(
		\frac{\sg_0}{d_R(\bA,\bB)+\sg_0}
	\right)^{12} -
	\left(
		\frac{\sg_0}{d_R(\bA,\bB)+\sg_0}
	\right)^{6} 
\right] = 
&
\nonumber \\
&
4 \epsilon_0 
\left[ 
	\left(
		\frac{\sg_0}{R- R \, F(\bA,\bB)^{-1/2} +\sg_0}
	\right)^{12} -
	\left(
		\frac{\sg_0}{R- R \, F(\bA,\bB)^{-1/2} +\sg_0}
	\right)^{6} 
\right],
&
\label{bigGB}
\end{eqnarray}
where $\sg_0$ has the  meaning of characteristic length and is responsible for the width of the potential well.  
The shape of the potential well is now more realistic because the distance $d_R(\bA,\bB)$ is a good approximation
to the distance $d(\bA,\bB)$. The potential (\ref{bigGB}) has the form of the GB potential without a  strength parameter and the 
PW range parameter  instead of the GOP one~\cite{GB}.

The depth of the  wells of the potential (\ref{bigGB}) however still equals one in any direction. One way to allow for variable depth of the potential minima  
is to use different shape matrices for the attractive and the repulsive  
part of the potential of Eq.(\ref{bigGB}) 
\begin{eqnarray}
\label{UABxi}
&
U(\bA_1,\bA_2,\bB_1,\bB_2) = 
&
\nonumber \\
&
4 \epsilon_0 
\left[ 
	\left(
		\frac{\sg_0}{R- R\,F_1(\bA_1,\bB_1)^{-1/2} +\sg_0}
	\right)^{12} -
	\left(
		\frac{\sg_0}{R- R\,F_2(\bA_2,\bB_2)^{-1/2} +\sg_0}
	\right)^{6} 
\right],
&
\end{eqnarray}
where the "repulsive ECF" $F_1(\bA_1,\bB_1)$ and the "attractive ECF" $F_2(\bA_2,\bB_2)$ 
are calculated using different shape matrices
\begin{equation}
\bA_1 = \sum_{i=1,2,3} a_{1i}^{-2} \bm{u}_i \otimes \bm{u}_i, \quad
\bB_1 = \sum_{i=1,2,3} b_{1i}^{-2} \bm{v}_i \otimes \bm{v}_i,
\end{equation}
and 
\begin{equation}
\bA_2 = \sum_{i=1,2,3} a_{2i}^{-2} \bm{u}_i \otimes \bm{u}_i \quad 
\bB_2 = \sum_{i=1,2,3} b_{2i}^{-2} \bm{v}_i \otimes \bm{v}_i \, .
\end{equation}

A justification for this approach can be given by inspecting the constant levels of the attractive 
and repulsive parts of a pair potential of two complex molecules, calculated as the sum of attractive and repulsive 
parts of the LJ potentials of spherical particles. It is easy to see that the constant levels $LJ_{12}$ and $LJ_{6}$ do have close but
different shapes. 

Although this introduces more parameters to the problem, the total  number of fitting parameters  is still  reduced compared 
to the GB potential. For example, the empirical exponents $\nu$ and $\mu$ present in the GB potential (see Appendix B) are excluded. Additionally, the  small deviation of the attractive shape matrices $\bA_2$ and $\bB_2$  
with respect to the repulsive ones $\bA_1$ and $\bB_1$ allows us in practise to use the solution of the "attractive" 
ECF problem- $\lm_{c2}$ and $\bx_{c2}$- as the initial conditions for the "repulsive" ECF problem, 
which reduces the number of iterations required for the calculation of the latter. 
Besides, when calculated separately, the repulsive ECF problem requires a  much smaller cut-off value.
As a result both ECF problems are solved only when  molecules are very close. 

To illustrate the form of the potential we consider two examples.
\subsection{Example 1}

\begin{figure}[htbp!]
\centering
\includegraphics{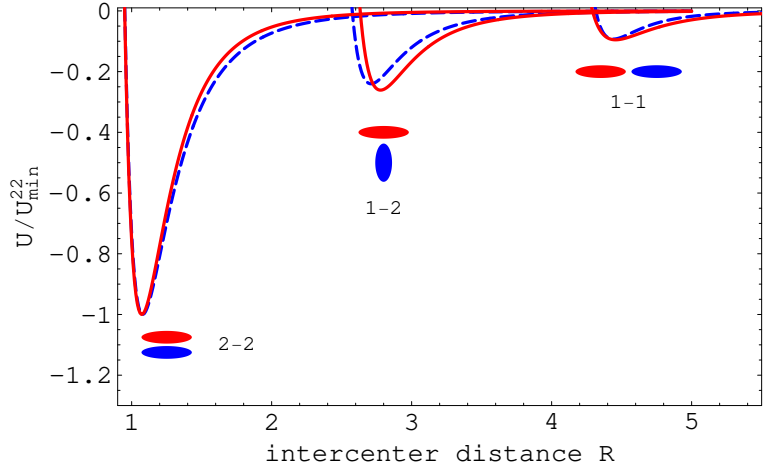}
\caption{ The resulting fitted  elliptic potential of Eq.~\ref{UABxi}  (red solid curves -) and the target atomistic LJ  potential (blue dashed curves -\,-\,\,) of two identical spheroids.  The original system consists of  6 
identical spherical LJ particles separated by a distance of $2/3$ and LJ parameters  $\sigma=1$, $\epsilon=1$.  The ``2-to-2'' or ``side-to-side''  minimum of such potential is $min_R U_{22} = 14.883$ which has been set to unity. Although only the values of  ``1-to-1'' and ``2-to-2'' potential minima  are fitted, the ``1-to-2'' or "side-to-end" configuration is also close to the target potential. The width parameter $\sigma_o$ is equal to unity. The semi-axes  of the repulsive matrices  $\bA_1$ and $\bB_1$ are 
$a_{11}=b_{11}=0.475 , a_{12}=b_{12}=2.025$, while the semi-axes of the attractive matrices $\bA_2$ and $\bB_2$ are given by 
$ a_{21}=b_{21}=0.475, a_{22}=b_{22}=1.875$.}
\label{FittedPot}
\end{figure}

As a simple test case, we consider the pair potential of two identical spheroids consisting of 6 LJ
spherical particles in a linear array \cite{GB}.  The resulting fitted potential and the target LJ  potential
are shown in  Fig. \ref{FittedPot}. Although only the values of  "side-to-side" and 
"end-to-end" potential minima  are fitted, the "side-to-end" configuration is also close to the target potential.  The ``2-2'' or ``side-to-side'' potential minimum of such potential is $min_R U_{22} = 14.883$ when the  LJ parameters are $\sigma=1$, $\epsilon=1$, and the spheres are separated by a distance of $2/3.$ and has been set to unity. Therefore a strength parameter $\epsilon = 14.883$
should be used in simulations.

For identical uniaxial particles there is the  possibility for further reduction of the number 
of the fitting parameters.  
The choice of "{\it i}-th" "attractive semi-axes" equal to "{\it i}-th" "repulsive semi-axes"
\begin{equation}
\label{aibi}
a_{1i} = b_{1i} = a_{2i} = b_{2i}
\end{equation}
leaves the corresponding "{\it i}-to-{\it i}" potential minimum equal to one as expected. Then the parameter $\epsilon_0$ in Eq. (\ref{bigGB}) 
 has the  meaning of the "{\it i}-to-{\it i}" potential minimum. As a result, the adjustment of the pair potential (\ref{bigGB}) for identical spheroids requires in total 
5 fitting parameters, while in the GB potential 8 parameters are used.

The same reduction can be used for unequal and biaxial particles as well, but then the fitting of the potential to 
a target potential become less flexible in the  representation of ratios of different potential minima.

\subsection{Example 2}

\begin{figure}[htbp!]
\centering
\includegraphics[width=8cm]{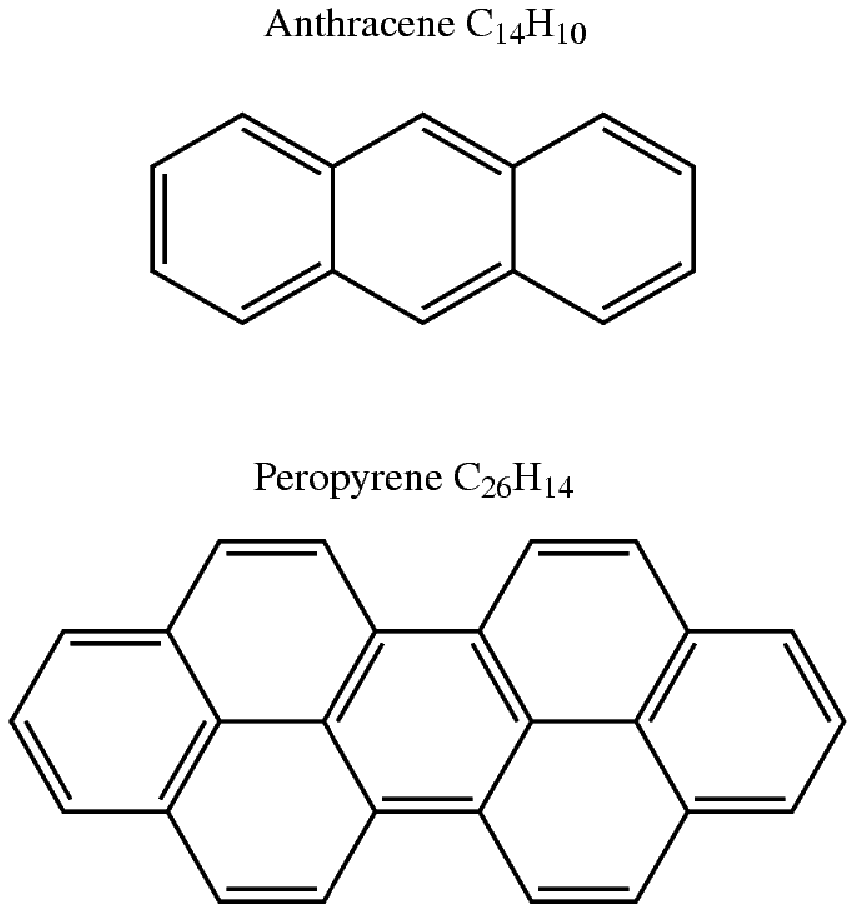}
\caption{The  test molecules of the   unequal  biaxial case: peropyrene, $C_{26}H_{14}$, and anthracene, $C_{14}H_{10}$.}
\label{PerAnthr}
\end{figure}

\begin{figure}[htbp!]
\centering
\includegraphics[width=13cm]{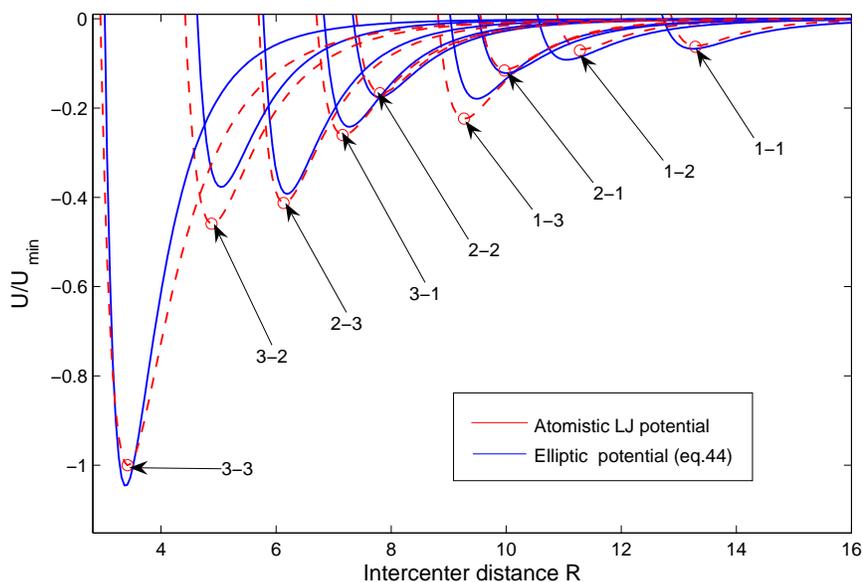}
\caption{"i-to-j" potential minima of the peropyrene-anthracene LJ atomic pair potential (red dashed lines -\,-\,-) fitted with the suggested potential of Eq.~\ref{UABxi} (blue solid curves-).  The fit was done with a weighted sum of squares difference which in this case gave more weight to the accuracy of the ``aligned'' configurations ``1-to-1'',''2-to-2'' and ``3-to-3''. This  choice  would depend on the particular application.  The following LJ parameters  were used for  hydrogen, H, and carbon, C,: $\sigma_{HH} = 2.4 \AA, \epsilon_{HH} = 0.02$ 
kcal/mol, $\sigma_{CC} = 3.4 \AA, \epsilon_{CC} = 0.15$ kcal/mol.
The resulting repulsive shape matrix elements for  $\bA_1$ and $\bB_1$ are 
$a_{11} = 7.024, a_{12} = 4.006,  a_{13} = 1.582$ and 
$b_{11} = 4.846, b_{12} = 2.841,  b_{13} = 1.511$, while  the attractive shape 
matrix elements for  $\bA_2$ and $\bB_2$ are given by  $a_{21} = 6.528,  a_{22} = 3.829,  a_{23} = 1.702, b_{21} = 4.267,  b_{22} = 2.405,  b_{23} = 1.394$, with width parameter $\sigma_0=2.921$ and  depth parameter $\epsilon = 25.543$.}
\label{PeropyrAntrFit}
\end{figure}

We consider  a more complicated example, namely  the  pair interaction of two 
non-identical biaxial molecules,  peropyrene $C_{26}H_{14}$ and anthracene $C_{14}H_{10}$ (Fig. \ref{PerAnthr}).
The peropyrene-to-anthracene pair potential is calculated as a sum of LJ interactions between all 
atoms of the molecules after a  geometry optimisation of the isolated molecules \cite{Gaussian}. Considering only cases where the "i-th" semi-axis of the particle "A" is  aligned with the "j-th" semi-axis of the particle "B", there 
are in total 9 "i-to-j" configurations which need to be represented.
 The objective function for the least-squares fitting was built as a
weighted sum of square differences of the  potential minimum value, the  position
of the root $U(\bA,\bB)=0$ and the  width of the potential well at half-depth
for each of the "i-to-j" relative orientations of the molecules. In this example, the relative weights were chosen to represent "aligned" configurations
"1-to-1", "2-to-2" and "3-to-3" more accurately then other "misaligned"
configurations. However, the particular choice of the weights can be made to
suite an application in mind.  The target potential was averaged over rotations of the molecules about
their inter-center vector $\bR$ for each "i-to-j" configuration.  Fig. (\ref{PeropyrAntrFit}) shows that  the potential (\ref{UABxi}) is able to reproduce, at least qualitatively, the complex interaction profile of  the peropyrene-anthracene pair potential. This is encouraging since $d_R$  is  more accurate at longer separations rather than at short but the wells around  the minima are still reasonably well represented. 

\section{Derivation of forces and torques}
\label{forces}

The dynamical evolution of the system via Molecular Dynamics (MD) simulations requires the calculation of forces and torques 
acting on the particles due to their interactions with all other particles. The suitability of a reduced representation 
potential for  MD simulations is mainly defined by the computing time spent on the calculation of forces and torques due to that potential. A numerical estimation of the derivatives of the potential is computationally expensive. Additionally,  approximate values of the derivatives is an extra source for numerical errors.  The  possibility of expressing those derivatives analytically is  a big advantage for a reduced representation potential.  

In this Section we derive analytic expressions for the forces and torques acting on the elliptic particles due to the 
suggested potential (\ref{UABxi}) that allow for efficient MD simulations to be performed. The resulting formulae are simple  due to the simple structure of the potential 
 (\ref{UABxi}) and the extremum properties of the ECF value (\ref{ECF_Val}).

%{\bf (*these notations are added*)}
The following notation will be used:
\begin{eqnarray}
&
U_{AB} = U(\bA_1,\bA_2,\bB_1,\bB_2),
\label{ShortNotationsUFG}
&
\\
&
F_i = F_i(\bA_i,\bB_i),
\nonumber
&
\\
&
\bX_{c} = \bX(\lm_c).
\nonumber
&
\end{eqnarray}

Given 
\begin{equation}
U_{AB} = 4 \epsilon_0 \left[ G_1^{-12} - G_2^{-6} \right] \, ,
\end{equation}
where 
\begin{equation}
G_i = ( R - R F_i^{-1/2} + \sg_0 ) / \sg_0,
\end{equation}
the force acting on particle "A" is
\begin{equation}
%&
\bm{F}_A=-\frac{d\, U_{AB}}{d \,\br} = 
%&
%\nonumber \\
%&
24 \epsilon_0
\left[ 
2 \, G_1^{-13} \,
  \frac{d\, G_1}{d \,\br}-
  G_2^{-7} \,
  \frac{d\, G_2}{d \,\br}
\right] \, , 
%&
\end{equation}
where
\begin{equation}
\label{dGdr}
\frac{d\, G_i}{d \,\br} =
\frac{1}{\sg_0}
\left[
\hat{\bR}(F_i^{-1/2}-1) +
\frac{R}{2} \, F_i^{-3/2} \frac{d \, F_i}{d \, \br} \, 
\right] \, ,
\end{equation}
and 
$$
\frac{d \, F_i}{d \, \br}=
- 2 \lm_{ci} \bA_i (\bx_{ci}-\br) = - \bX_{ci}\, ,
$$
with $\lm_{ci}$, the contact parameter, and $\bx_{ci}$, the contact point of the ECF task 
(Eqs \ref{ECF_Val}) - (\ref{Xl}) with matrices $\bA_i$ and $\bB_i$.  The derivatives of the ECF, $\frac{d \, F_i}{d \, \br}$,  are  given in Appendix C. 

It can be shown that 
$$
\bm{F}_B = - \bm{F}_A 
$$
and so  the force conservation equation is fulfilled. Similarly, the torque acting on particle "A" is 
\begin{eqnarray}
&
\bm{T}_A= - \frac{24 R \epsilon_0}{\sg_0} 
[
2 \, G_1^{-13}  F_1^{-3/2} 
  (\bx_{c1}-\br) \times \bX_{c1}  -
&
\nonumber \\
&
G_2^{-7} F_2^{-3/2} 
  (\bx_{c2}-\br) \times \bX_{c2}
]\, .
& 
\end{eqnarray}

The torque conservation equation (\cite{Forces}) is also fulfilled 
$$
\bm{T}_A +\bm{T}_B + \bR \times \bm{F}_B = 0 \, .
$$

\section{Discussion} 
\label{discussion}
Anisotropic particles are of interest for interrogating fundamental questions such as packing\cite{Torquato03} as well as for applications in biological and nanoscale systems. Unlike spherically symmetric 
particles where the distance of closest approach is immediately given 
by their inter-center distance, this is no longer true in anisotropic systems. Obtaining the distance of closest approach as a function of orientation for any size and shape is a difficult task. Additionally, the  potentials must  
capture the geometric and energetic anisotropies of the system consistently.

 In this work,  we have considered in detail the geometry of the ECF  as an approach measure of ellipsoid  pair particles. We have shown that the  
 directional distance of closest approach $d_R(\bA,\bB)$, measured along the inter-center vector,  can be  derived 
directly from the ECF. Like the true distance of closest approach of two ellipsoids, $d(\bA,\bB)$,   $d_R(\bA,\bB)$ is  zero when the ellipsoids are in 
contact.  Unlike $d(\bA,\bB$, which cannot be used in its  regular form (\ref{dMinAB})-(\ref{ConstrMinAB}) to characterize any overlap of the ellipsoids,  $d_R(\bA,\bB)$ is also a measure of their overlap.   As a result, "soft" elliptic potentials can also be built.  The geometrical meaning of the Perram-Wertheim approach parameter $\sg_{PW}(\bA,\bB)$ is clarified as the shortest directional distance of closest approach along the inter-particle vector $\bR.$ 
Note that   the volume of elliptic 
particles is preserved within the ECF approach.

 Since this approach is valid for any two ellipsoids, an elliptic potential is suggested which can be used for  the modelling of any  mixture of elliptic and spherical particles.   Examples  of the suggested fitted potential to pair potentials of different biaxial molecules has shown it is able to reproduce, at least qualitatively, complex interaction profiles. Additionally, it correctly reduces to isotropic interactions in both shape and magnitude at long distances. For questions  such as how  self-assembly emerges,  it may be  more important to sacrifice quantitative accuracy at short  range in order to obtain a consistent long range interaction that does not overestimate  the range of the potential and hence impose assembly artificially.

The structure of the potential leads to fewer   fitting parameters than the GB potential. Analytic expressions for forces and torques acting on the particles due to the suggested potential are derived, which make it amenable to MD simulations.  However,  an iterative solution of the Eq. (\ref{lc_EQ}) is required for each pair of molecules on each 
integration step during the MD simulation. Although this can be done efficiently, the GB potential is computationally more effective from this point of view. On the other hand,   we have  shown in Sec.~\ref{comparison}, that the GB potential leads to deviations of the shape and of the volume
of interacting molecules whenever the semi-axes of the ellipsoids are different or misaligned.  The described deviations of the volume of the elliptic particles lead  to an additional artificial  "ordering" force which might be desirable 
in modelling of bulk equilibrium phases, such as liquid crystals, but may  not be as  
reasonable when considering dynamical properties  in  mixtures of dissimilar  nanoscale particles whose properties are strongly dependent on their size and shape. From this point of view, there is a "trade-off"   between computational efficiency and  accuracy in the  treatment 
of the shape and volume of the molecules.  The PW approach parameter and the suggested elliptic pair potential  
are one of possible ways to resolve this. 

Another weakness  of the  potential of Eq. (\ref{UABxi}), in common with the ECP  potential,  has already  been  
discussed in \cite{EveraersEjtehadi2004}. For instance, in the case of the interaction of two uniaxial elliptic particles 
there are 
infinite number of possible "side-to-side" configurations corresponding to rotations of particles about their 
inter-particle vector $\bR$.
The extremities are the "parallel" configuration, when the main axis of the particles are aligned and 
the "crossover", when the main axes of the ellipsoids are orthogonal. 
For molecules consisting of a  linear array of spherical LJ particles these two configurations
obviously have different values of the potential minima, but they are the same in the  ECP and the suggested 
potential  as well. 
To take into account these effects one should include into the potential a  dependence on the local curvature tensors of 
surfaces $\cA(\bx)=const$ and $\cB(\bx)=const$ at the contact point $\bx_c$ or at sub-contact points  
$\bx_a$ and $\bx_b$.

A problem  in connection with this path is that the curvature of the surface of an ellipsoid in most 
cases is not  the best representation of  the shape of a real molecule. The fitting of an "i-to-j" configuration of a 
molecule with an elliptic potential can only reproduce the characteristic length in its main directions but not the total shape.
 A more sophisticated model of molecular shape is needed to take these effects into account. The ECF approach in its general form of Eq. (\ref{sMinAB}) can be extended to the case of a general 
convex body\cite{future}. Then it probably should be called a "Convex Contact Function approach". The approach keeps the formulas for forces and torques in almost the same form. 
The computational efficiency of the Convex Contact Function extension of the ECF depends upon the description of the surfaces of convex bodies. Work in this direction is in progress.

\section*{Acknowledgements} 
We thank John Perram for useful discussions and for providing Ref. \cite{JohnUnPblshd}. LP thanks Lula Rosso for providing the  optimised geometries of the  example. 
This research has been supported by GlaxoSmithKline.

\section*{Appendix A. The relation between the two definitions of the ECF}

%%{\bf (*the beginning is rewritten*)}
The definition (\ref{sMinAB})-(\ref{ECF_Val}) of the ECF which is used in this article 
first appeared in the unpublished work \cite{JohnUnPblshd} of Perram. 
In this Appendix we show that it is  equivalent to the original definition given in  Ref. \cite{John2}. 

The inter-particle vector $\bR = (\bs-\br)$ along the curve $\bx(\lm)$ can be expressed as
\begin{eqnarray}
&
\bR = (\bx(\lm)-\br)-(\bx(\lm)-\bs) =
\label{RxlX}
&
\\
&
\lm^{-1}\bA^{-1} \cdot \bX(\lm) + (1-\lm)^{-1} \bB^{-1} \cdot \bX(\lm).
\nonumber
&
\end{eqnarray}
The vector $\bX(\lm)$ itself can be found from (\ref{RxlX}) as a solution to  the following linear equation
\begin{equation}
\label{XR}
\bX(\lm) \cdot 
\left\lbrace 
\lm^{-1} \bA^{-1} + (1-\lm)^{-1} \bB^{-1}
\right\rbrace = \bR \, .
\end{equation}
Using Eq (\ref{BigXl}) for the scaled gradient vector $\bX(\lm)$, 
the value of the quadratic form $\cS(\bx(\lm),\lm)$ now becomes
\begin{eqnarray}
\label{lxcX}
&
\cS(\bx(\lm),\lm) = 
\lm \cA(\bx(\lm)) + (1-\lm) \cB(\bx(\lm))=
&
\nonumber \\
&
(\bx(\lm) - \br)^T \cdot \bX(\lm) - (\bx(\lm) - \bs)^T \cdot \bX(\lm) = 
& 
\nonumber \\
&
\bX^T(\lm) \cdot
\left\lbrace 
\lm^{-1} \bA^{-1} + (1-\lm)^{-1} \bB^{-1}
\right\rbrace \cdot \bX(\lm) \, .
&
\end{eqnarray}
The quadratic form in the expression above  has the same matrix as the linear equation (\ref{XR}). 
Substituting Eq. (\ref{XR}) into Eq. (\ref{lxcX}), we obtain  
\begin{eqnarray}
&
\cS(\bx(\lm),\lm) =
\bX^T(\lm) \cdot
\left\lbrace 
\lm^{-1} \bA^{-1} + (1-\lm)^{-1} \bB^{-1}
\right\rbrace \cdot \bX(\lm) = 
&
\nonumber \\
&
\lm (1-\lm) \bm{\hat{R}}^T \cdot
\left\lbrace 
(1-\lm) \bA^{-1} + \lm \bB^{-1}
\right\rbrace^{-1} \cdot \bm{\hat{R}} R^2 \, .
&
\label{SRhat}
\end{eqnarray}
This is 
the original definition of the quadratic form (\ref{SRhat}) given in \cite{John1} and \cite{John2}.

\section*{Appendix B:  The  Gay-Berne  potential}

The GB potential for identical uniaxial particles 
(see \cite{GB},\cite{GBE}, \cite{GBE1}) has the following form: 
\begin{equation}
\label{GBEeqn}
U_{GB} = 
4 \epsilon_0 
\epsilon_1^{\nu}( \bA , \bB \, ) \,
\epsilon_2^{\mu}( \bm{E}_1 , \bm{E}_2, \hat{\bR} \,) \,
(\eta^{12} - \eta^{6}) \, , 
\end{equation}
\begin{equation}
\eta = \frac{\sg_0} 
{( R - \sg_{BP} ( \bA , \bB, \hat{\bR} \, )  +\sg_0 )} \, ,
\nonumber
\end{equation}
where the Berne-Pechukas range parameter $\sg_{BP} ( \bA , \bB, \hat{\bR} \, )$ is given by 
\begin{equation}
\sg_{BP} ( \bA , \bB, \hat{\bR} \, ) = 
\left[ 
\frac{1}{2} 
	\left( 
	\hat{\bR}^T \cdot  
		\left\lbrace 
		\bA^{-1}+\bB^{-1} 
		\right\rbrace^{-1} 
	\cdot \hat{\bR}
	\right) 
\right]^{-1/2} \,
\label{sGOPdef}
\end{equation}
%{\bf and eo, e1,m, so...define all symbols}
The strength parameter $\epsilon_2 ( \bm{E}_1 , \bm{E}_2, \hat{\bR} \, )$ has the form of the 
square of the range parameter (\ref{sGOPdef}) calculated with different shape matrices.
\begin{equation}
\epsilon_2 ( \bm{E}_1 , \bm{E}_2, \hat{\bR} \, ) =
\frac{1}{2} 
	\left( 
	\hat{\bR}^T \cdot  
		\left\lbrace 
		\bm{E}_1+\bm{E}_2 
		\right\rbrace^{-1} 
	\cdot \hat{\bR}
	\right).
\label{E2}
\end{equation}
The matrices $\bm{E}_1$ and $\bm{E}_2$ are defined as
\begin{equation}
\bm{E}_1 = \sum_{i=1,2,3} \epsilon_{1i}^{-1/\mu} \bm{u}_i \otimes \bm{u}_i, \quad
\bm{E}_2 = \sum_{i=1,2,3} \epsilon_{2i}^{-1/\mu} \bm{v}_i \otimes \bm{v}_i \, , 
\end{equation}
where $\bm{u}_i, i=1,2,3$ are unit vectors along the semi-axes of ellipsoid "A" and vectors 
$\bm{v}_i, i=1,2,3$ are unit vectors along the semi-axes of ellipsoid "B".
The parameters $\epsilon_{1i}$ and $\epsilon_{2i}$ are responsible for the potential minima of side-to-side, side-to-end and 
end-to-end configurations. 

Using the expression (\ref{SRhat}) for the quadratic form  $\cS(\bx(\lm),\lm)$,
the BP range parameter can be found by substituting $\lm=1/2$.
\begin{eqnarray}
&
\sg_{BP}(\bA,\bB) = 
\frac{R}{\sqrt{\cS(\bx(1/2),1/2)} } \, .
& 
\end{eqnarray}

\section*{Appendix C: Derivatives of the  ECF}

The derivative of the ECF, $F(\bA,\bB)$ (Eq. \ref{}), with respect to position, $\br$, for ellipsoid  "A" is given by 
\begin{equation}
%&
\frac{d F(\bA,\bB)}{d \br}  =  
\frac{\partial \cS(\bx_c,\lm_c)}{\partial \br} + 
\label{dFdrFull}
%&
%\\
%&
\cS'_{\lm}(\bx_c,\lm_c) \, \frac{d \lm_c}{d \br} +
\nabla_{\bx} \cS(\bx_c,\lm_c) \, \frac{d \bx(\lm_c)}{d \br}
%\nonumber
%&
\end{equation}
where we have used the notation introduced in Eq.~( \ref{ShortNotationsUFG}).
The derivative of Eq. \ref{dFdrFull} is evaluated at the contact parameter $\lm=\lm_c$. This greatly simplifies the expression when we notice that 
the term $\nabla_{\bx}\cS(\bx(\lm),\lm)$ vanishes for any point on the curve $\bx(\lm)$ due to Eq. (\ref{dFdl}) and 
the term $\cS'_{\lm}(\bx(\lm),\lm)$ vanishes at the extremum point due to Eq. (\ref{dFdlsmall}). 
As a result, we are left with 
\begin{equation}
%&
\frac{d F_i}{d \br}  =  
\frac{\partial \cS(\bx(\lm),\lm)}{\partial \br}  = 
\label{dFdr}
%&
%\\
%&
- 2 \, \lm_{ci} \bA_i \cdot (\bx_{ci}-\br) = - \bX_{ci} \, .
%\nonumber
%&
\end{equation}
Similarly,  for  particle "B"  
\begin{equation}
\frac{d F_i}{d \bs}  =  
- 2 (1-\lm_{ci}) \bB_i \cdot (\bx_{ci}-\bs) = \bX_{ci}
\end{equation}

 The torque $T_{Aj}$ about an axis $\be_j$  and rotation angle  $\psi_j$ 
 due to the potential (\ref{UABxi}) is 
\begin{eqnarray}
&
T_{Aj} = -\frac{d \, U_{AB}}{d\,\psi_j} 
= 
-\sum\limits_{i=1}^2 \frac{d\,U_{AB}}{d \, F_i}\,\frac{d\, F_i}{d\,\psi_j} \, ,
&
\end{eqnarray}
where
\begin{eqnarray}
\frac{d U_{AB}}{d F_1} 
& 
= 
& 
- 4 \epsilon_0 
\left[
12 G_1^{-13}  F_1^{-3/2} {R \over 2 \sg_0} 
\right] \, ,
\\
\frac{d U_{AB}}{d F_2} 
& 
= 
&
4 \epsilon_0 
\left[ 
6  G_2^{-7} F_2^{-3/2} {R \over 2 \sg_0 }
\right] \, ,
\end{eqnarray}
where $G_i$ is given in Eq.~(35) (see also Eq.~(37)).
The derivative $d \, F_i/d \, \psi_j$, similarly  to Eqs (\ref{dFdrFull})-(\ref{dFdr}), is
\begin{equation}
\frac{d F_i}{d \psi_j}  = 
\lm_{ci} (\bx_{ci}-\br)^T \cdot \frac{d \bA_i}{d \psi_j} \cdot (\bx_{ci}-\br) \, .
\label{dFdPsiJ}
\end{equation}
To calculate $d \bA/d \psi_j$, we write $\bA = \bm{P} \cdot \bA_0 \cdot \bm{P}^T$, 
where $\bm{P}=\bm{P}(\psi)$ is the rotation matrix about the axis $\be$ by the angle $\psi$ given,  in dyadic form, by
\begin{equation}
\bm{P}(\psi) = 
\be \otimes \be + 
( 1-\cos(\psi) ) \bm{E} +
\sin(\psi) \, \be \times \bm{E} \, .
\end{equation}
and $\bA_0$ is a diagonal matrix of the quadratic form $\cA(\bx)$ in the "body reference frame" attached 
to the semi-axes of the ellipsoid "A". Using the well-known expression $\bm{P}'_{\psi}= \be \times \bm{P}$ 
for the derivative of the rotation matrix $\bm{P}$, the derivative $d \bA/d \psi_j$ can now be expressed as
\begin{eqnarray}
&
\frac{d \bA}{d \psi_j} = 
\frac{d \bm{P}}{d \psi_j} \cdot \bA_0 \cdot \bm{P}^T +
\bm{P} \cdot \bA_0 \cdot (\frac{d \bm{P}}{d \psi_j})^T = 
&
\nonumber \\
&
\be_j \times \bm{P} \cdot \bA_0 \cdot \bm{P}^T + 
\bm{P} \cdot \bA_0 \cdot (  \be_j \times \bm{P} )^T =
&
\nonumber \\
&
\be_j \times \bm{P} \cdot \bA_0 \cdot \bm{P}^T - 
\bm{P} \cdot \bA_0 \cdot \bm{P}^T \times \be_j  =
&
\nonumber \\
&
\be_j \times \bA - \bA \times \be_j = 2 \, \be_j \times \bA.
&
\end{eqnarray}
The last equality follows from the  symmetry of the matrix $\bA_0$. Using the tensor equality $\bm{a} \cdot (\bm{b} \times \bm{C} ) = (\bm{a} \times \bm{b}) \cdot \bm{C}$, where 
$\bm{a}$ and $\bm{b}$ are vectors and $\bm{C}$ is a second rank tensor, Eq. (\ref{dFdPsiJ}) becomes 
\begin{eqnarray}
&
\frac{d F_i}{d \psi_j}  = 
2 \lm_{ci} (\bx_{ci}-\br)^T \cdot (\be_j \times \bA_i) \cdot (\bx_{ci}-\br) =
&
\\
&
2 \lm_{ci} ((\bx_{ci}-\br)^T \times \be_j) \cdot \bA_i \cdot (\bx_{ci}-\br) =
&
\nonumber
\\
&
( (\bx_{ci}-\br) \times \be_j ) \cdot \bX_{ci} =
( \bX_{ci} \times (\bx_{ci}-\br) ) \cdot  \be_j \, ,
&
\nonumber
\end{eqnarray}
where the last operation is just a cyclic permutation of vectors in the calculation of the volume 
of a parallelogram built  by the  three vectors $(\bx_{ci}-\br)$, $\be_j$ and $\bX_{ci}$.

Substituting Eq. 73 into Eq. 67, the total torque vector $\bm{T}_{A}$ acting on the particle "A", 
$\bm{T}_{A}=\sum\limits_{j=1}^3 T_{Aj} \be_j$
\begin{eqnarray}
&
\nonumber \\
& 
-\sum\limits_{i=1}^2 \frac{d  U_{AB}}{d F_i} 
\left[ 
\sum\limits_{j=1}^3 ( \bX_{ci} \times (\bx_{ci}-\br) ) \cdot \be_j \be_j 
\right] =
&
\nonumber \\
&
-\sum\limits_{i=1}^2 \frac{d  U_{AB}}{d F_i} 
\left[  
( \bX_{ci} \times (\bx_{ci}-\br) ) \cdot \bm{E} ) 
\right] =
&
\nonumber \\
&
\sum\limits_{i=1}^2 \frac{d U_{AB}}{d F_i} 
\left[  
(\bx_{ci}-\br) \times \bX_{ci} 
\right] \, ,
&
\end{eqnarray}
where $\bm{E}$ is the unit tensor
$
\bm{E} = \sum\limits_{j=1}^3 \be_j \otimes \be_j.
$

\bibliography{MyBib}

%
%\begin{figure}[htbp]
%\centering
%\includegraphics{FigPlot2}
%\caption{Here and further without loss of generality the concepts behind ECF approach will be illustrated by 2D case.  
%Original ellipsoids "A" (red thick closed curve) and "B" (blue thick closed curve) are considered in the local
%reference frame attached to the centre $\br$ of the ellipsoid "A" aligned along its semi-axis.  
%The surface $\cA(\bx)=\cB(\bx)$ is shown with the thin black line. The curve $\bx(\lm)$ is shown 
%with the thin green line. The contact point $\bx_c$ is the intersection of $\bx(\lm)$ with 
%$\cA(\bx)=\cB(\bx)$}
%\end{figure}

\newpage

%\begin{figure}[htbp]
Fig.1: The ellipsoid particles   "A" and "B", given by $\cA(\bx)=1$ and $\cB(\bx)=1$, are centered on $\br$ and $\bs$ respectively. We consider the local
reference frame attached to the centre  of  ellipsoid "A" aligned along its semi-axis.  We illustrate here, without loss of generality, the concepts behind the ECF approach  in the 2D case.  The original ellipsoids  are scaled up
(or down) by $\sqrt{F(A,B)}$ until they touch each other tangentially at the contact point $\bx_c$. The contact point $\bx_c$ is the intersection of the curve $\bx(\lm)$ (green line)  with the surface 
$\cA(\bx)=\cB(\bx)$ (black line).
The value of the ECF is illustrated by the scaled ellipsoids $\cA(\bx_c)=F(\bA,\bB)$ and
$\cB(\bx_c)=F(\bA,\bB)$.
%\label{TwoScaledElls}
%\label{TwoEllsAndXC}
%\end{figure}
%\pagebreak
%\begin{figure}[htbp]

Fig.2: The  directional distance of closest approach, $d_R(A,B)$,  is the minimum distance  between the surfaces of two ellipsoids   that is parallel to the inter-center vector R. It  is given by the  sub-contact points $\bx_a$ and $\bx_b$, which are on  the surface of ellipsoids ``A'' and ``B'', respectively. It is a good approximation from above to the true contact distance.    
The value of the PW 
approach parameter $\sg_{PW}(\bA,\bB)$
can be  understood by the  geometry of the triangle of points $\br$, $\bs$ and $\bx_c$, as the length of the vector $\bR -(\bx_a$- $\bx_b)$ as soon as $(\bx_a$- $\bx_b)$ becomes parallel to $\bR$ 
%\label{ExtraSubPoints}
%\end{figure}
%\pagebreak
%\begin{figure}[htbp]

Fig.3: The distance $d_n(A,B)$  is defined as the distance between two parallel planes $\alpha_a$ and $\alpha_b$ which are tangent  at any point $\tilde{\bx_a}$ and $\tilde{\bx_b}$ on the surface of  ellipsoids ``A'' and ``B'', respectively. It approaches the true distance of closest approach  from below. The maximum distance $d_n(A,B)£$ coincides with  the true distance of closest approach. This can be understood from a mechanical point of view: if the planes are kept apart by  a constant force ${\bf F}$, then the equilibrium point reached   corresponds to the maximum distance between planes, which is also the distance of closest approach.
%\label{MaxDNDist}
%\end{figure}
%\pagebreak

%\begin{figure}[htbp]
Fig.4: A closer look at the sub-contact points $\bx_a$ and $\bx_b$, which define the directional distance of closest approach $d_R$. The planes $\alpha_a$ and 
$\alpha_b$ are  tangent to the ellipsoids $\cA(\bx)=1$ and $\cB(\bx)=1$ respectively at these points (shown with 
purple lines). The distance $d_n$ approaches  the true distance of closest approach  from below, while $d_R$ approaches it  from above.
%\label{Planes}
%\end{figure}

%\begin{figure}[htbp]
Fig.5:  $\cS(\bx(\lm),\lm)$ is shown as a function of the parameter $\lambda$  for the ``2-to-2'' ("side-to-side"),
``1-to-1'' ("end-to-end") and ``1-to-2'' ("side-to-end") relative orientations of two identical ellipsoids. The  ECF value, ${F_{ij}(A,B)}$, is the maximum of the function $S$ and gives the contact parameter $\lambda_c$ for  each orientation.  The approximation of 
$\lm_c$ by $1/2$ holds for symmetric configurations of identical ellipsoids such as ``2-to-2''  and ``1-to-1'',      but  fails in the asymmetric ``1-to-2''    configuration. This is expected to hold  for asymmetric configurations in general, which includes any configuration of non-identical particles.
%\label{SFunctionPlot}
%\end{figure}
%\pagebreak

%\begin{figure}[htbp]
Fig.6: The resulting fitted  elliptic potential of Eq.~\ref{UABxi}  (red solid curves -) and the target atomistic LJ  potential (blue dashed curves -\,-\,\,) of two identical spheroids.  The original system consists of  6 
identical spherical LJ particles separated by a distance of $2/3$ and LJ parameters  $\sigma=1$, $\epsilon=1$.  The ``2-to-2'' or ``side-to-side''  minimum of such potential is $min_R U_{22} = 14.883$ which has been set to unity. Although only the values of  ``1-to-1'' and ``2-to-2'' potential minima  are fitted, the ``1-to-2'' or "side-to-end" configuration is also close to the target potential. The width parameter $\sigma_o$ is equal to unity. The semi-axes  of the repulsive matrices  $\bA_1$ and $\bB_1$ are 
$a_{11}=b_{11}=0.475 , a_{12}=b_{12}=2.025$, while the semi-axes of the attractive matrices $\bA_2$ and $\bB_2$ are given by 
$ a_{21}=b_{21}=0.475, a_{22}=b_{22}=1.875$.
%\label{FittedPot}
%\end{figure}
%\pagebreak
%\begin{figure}[htbp]

Fig.7: The  test molecules of the   unequal  biaxial case: peropyrene, $C_{26}H_{14}$, and anthracene, $C_{14}H_{10}$.

%\label{PerAnthr}
%\end{figure}
%\pagebreak

%\begin{figure}[htbp]
Fig.8: "i-to-j" potential minima of the peropyrene-anthracene LJ atomic pair potential (red dashed lines -\,-\,-) fitted with the suggested potential of Eq.~\ref{UABxi} (blue solid curves-).  The fit was done with a weighted sum of squares difference which in this case gave more weight to the accuracy of the ``aligned'' configurations ``1-to-1'',''2-to-2'' and ``3-to-3''. This  choice  would depend on the particular application.  The following LJ parameters  were used for  hydrogen, H, and carbon, C,: $\sigma_{HH} = 2.4 \AA, \epsilon_{HH} = 0.02$ 
kcal/mol, $\sigma_{CC} = 3.4 \AA, \epsilon_{CC} = 0.15$ kcal/mol.
The resulting repulsive shape matrix elements for  $\bA_1$ and $\bB_1$ are 
$a_{11} = 7.024, a_{12} = 4.006,  a_{13} = 1.582$ and 
$b_{11} = 4.846, b_{12} = 2.841,  b_{13} = 1.511$, while  the attractive shape 
matrix elements for  $\bA_2$ and $\bB_2$ are given by  $a_{21} = 6.528,  a_{22} = 3.829,  a_{23} = 1.702, b_{21} = 4.267,  b_{22} = 2.405,  b_{23} = 1.394$, with width parameter $\sigma_0=2.921$ and  depth parameter $\epsilon = 25.543$.
%\label{PeropyrAntrFit}
%\end{figure}

\newpage

\newpage

\newpage

\newpage

\newpage

\newpage

\newpage

%\pagebreak
\newpage

%\pagebreak

\end{document}